\documentclass[english,twocolumn]{revtex4-1}
\usepackage[T1]{fontenc}
\usepackage[latin1]{inputenc}
\usepackage{babel}
\usepackage{epsfig}
\usepackage{amsmath}
\usepackage{amssymb}
\usepackage{epstopdf}
\usepackage{overpic}
\usepackage{color}
\usepackage{hyperref}

\hypersetup{colorlinks, citecolor = {blue}, linkcolor=black, pdftitle = {C. Mulhern, D. Hennig, and A.D. Burbanks: The coupled dynamics of two particles with different limit sets},
}

\begin{document}

\title
{\bf The coupled dynamics of two particles with different limit sets}
\author{C. Mulhern$^{1,2}$*, D. Hennig$^2$, and A.D. Burbanks$^2$}
\footnotetext{The text of the footnote goes here.}
\medskip
\medskip
\medskip
\affiliation{$^1$Max Planck Institute for the Physics of Complex System, 01187 Dresden, Germany}
\affiliation{$^2$Department of Mathematics, University of Portsmouth, Portsmouth, PO1 3HF, UK}

\email[Corresponding author: ]{mulhern@pks.mpg.de}

\begin{abstract}
\noindent 
We consider a system of two coupled particles evolving in a periodic and spatially symmetric
potential under the influence of external driving and damping. The particles are driven individually
in such a way that in the uncoupled regime, one particle evolves on a chaotic attractor, while
the other evolves on regular periodic attractors. Notably only the latter supports coherent particle transport. The influence of the coupling between the particles is
explored, and in particular how it relates to the emergence of a directed current. We show that  increasing the (weak) coupling
strength subdues the current in a process, which in phase-space, is related to a merging crisis of attractors 
forming one large chaotic attractor in phase-space. Further, we demonstrate
that complete current suppression coincides with a chaos-hyperchaos transition. 
\end{abstract}

\maketitle
\section{Introduction}
\noindent
Nonlinear transport processes continue to be of vital importance to the understanding of many physical systems.
In particular, the transport of particles in symmetric and periodic potential landscapes has attracted considerable
interest ~\cite{Acevedo}--\cite{HBO2}. This ubiquitous class of potentials lends itself to a vast number
of applications including Josephson junctions \cite{Mea}, charge density waves, nano engines \cite{AH}, and
transport in biological systems \cite{review}. More recently, a study has examined features of the possible interaction of a Josephson junction with axionic dark matter \cite{Beck}. They note the similarity between the equations of motion of axions and of Joesphson junctions (indeed the equations of motion used in the present study are strikingly similar to the coupled system consisting of an axion and a Josephson Junction) and highlight that their approach of coupling the two could lead to the future development of axionic dark matter detection.

The two-particle system (dimer) is widely studied in the context of these models precisely because it is
the smallest form consisting of coupled units and gives insight into what happens in systems of larger scale.
The formation of a bond between particles (monomers), often referred to as \emph{dimerisation}, plays an important role in transport processes.
This bond, on the one hand, can allow particles trapped in a potential well to escape to some
attracting domain, which individually they would be unable to do \cite{fugmann}. On the other hand,
this bond can cause erratic movement of the particles making up the dimer, and thus exclude the possibility
of a coherent escape process. Renewed interest in two-particle systems has also come about due to recent work
on Bose-Einstein condensates \cite{smerzi}.

Another interesting feature of two-particle systems, as compared to single-particle systems, is that a new
type of motion is now possible -- namely, hyperchaos \cite{ott}. Hyperchaos occurs when more than one of the
system's Lyapunov exponents is positive. This type of motion is characterised by the chaotic diffusion of
trajectories in two (or more) different phase-space directions. Some have related the chaos-hyperchaos transition to a change in the stability of an infinite set of unstable periodic orbits which, when a critical transition point is crossed, act as repellers  \cite{yanchuk}.  

In this paper we explore the deterministic nonlinear dynamics of two driven and damped coupled particles evolving
in a periodic and symmetric potential landscape. Individually, the dynamics of the particles is such that one
particle evolves on a chaotic attractor, while the other evolves on a periodic attractor. The dynamics of a similar
system, where the driving force for each particle is equal, has been studied recently \cite{chaos}. It was shown that coordinated energy exchange between the particles takes place such that they are able to surmount consecutive barriers of the periodic potential resulting in coherent transport. Similarly, in the present study
we are interested in the dynamics when the particles are coupled to one another. This coupling induces a competition
between two dynamical regimes. Therefore, the strength of the coupling will play a key role. In particular, in the scope of weak coupling we wish
to establish a critical coupling strength, beyond which directed particle transport is excluded. In previous studies
of one-particle systems, a bifurcation point, triggering a current reversal or a current suppression, has been readily
obtained numerically \cite{Mateos1,Mateos2}. However, the higher dimensional nature of the system explored in this paper
makes finding such a bifurcation point more difficult. To this end, we calculate the Lyapunov exponents for the system
and show that the emergence of a vanishingly small current coincides with the transition of the second largest Lyapunov
exponent to zero and subsequently becoming positive. Further, by examining three-dimensional projections of the phase-space
we will show that attractors, providing directed transport in opposite directions for low coupling strengths, expand
and finally merge as a result of a merging crisis as the coupling strength, playing the role of the bifurcation parameter,
 is increased. This coalescence of the attractors means that transporting
channels become suppressed with the particle dynamics being captured on a single chaotic attractor preventing the occurrence of a current at all. 

The paper is organised as follows: In the next section we introduce the system and discuss the individual particle dynamics; in section \ref{sect:current} we statistically assess the mean velocity which we relate to the system's Lyapunov spectrum and then later in section \ref{sect:pss} to the structures in phase-space; we conclude in section \ref{sect:summary} with a summary.       
    
\section{System}
\noindent
Let us consider a system of two coupled particles, each evolving in a spatially 
periodic
potential with spatial period $L=1$,
subjected to damping and periodic driving. The system is modelled by the following set of coupled ordinary differential equations

\begin{eqnarray}
\ddot{q}_1&=&-\sin(2\pi q_1)-\gamma \dot{q}_1+\,h_1\sin(\Omega\, t+\theta_0)-\kappa(q_1-q_2)\,\,,\label{eq:q1}\\
\ddot{q}_2&=&-\sin(2\pi q_2)-\gamma \dot{q}_2+h_2\sin(\Omega\, t+\theta_0)+\kappa (q_1-q_2)\,\,,\label{eq:q2}
\end{eqnarray}
\noindent
where the parameters $\gamma\geq0$, $h_{1,2}\geq0$, and $\kappa\geq0$ regulate the strength of damping,
driving, and coupling respectively. The driving frequency, common to both particles, is given by the
parameter $\Omega\geq0$. Throughout the paper the phase of the external driving field is $\theta_0=0$. 
 Quite often in our study $\kappa$ will be used as a control parameter.
Note that the exchange symmetry $(\dot{q}_1,q_1) \longleftrightarrow (\dot{q}_2, q_2)$
is not present in this model, except when $h_1=h_2$.

\begin{figure*}[ht!]
\includegraphics[height=7.2cm, width=8.4cm]{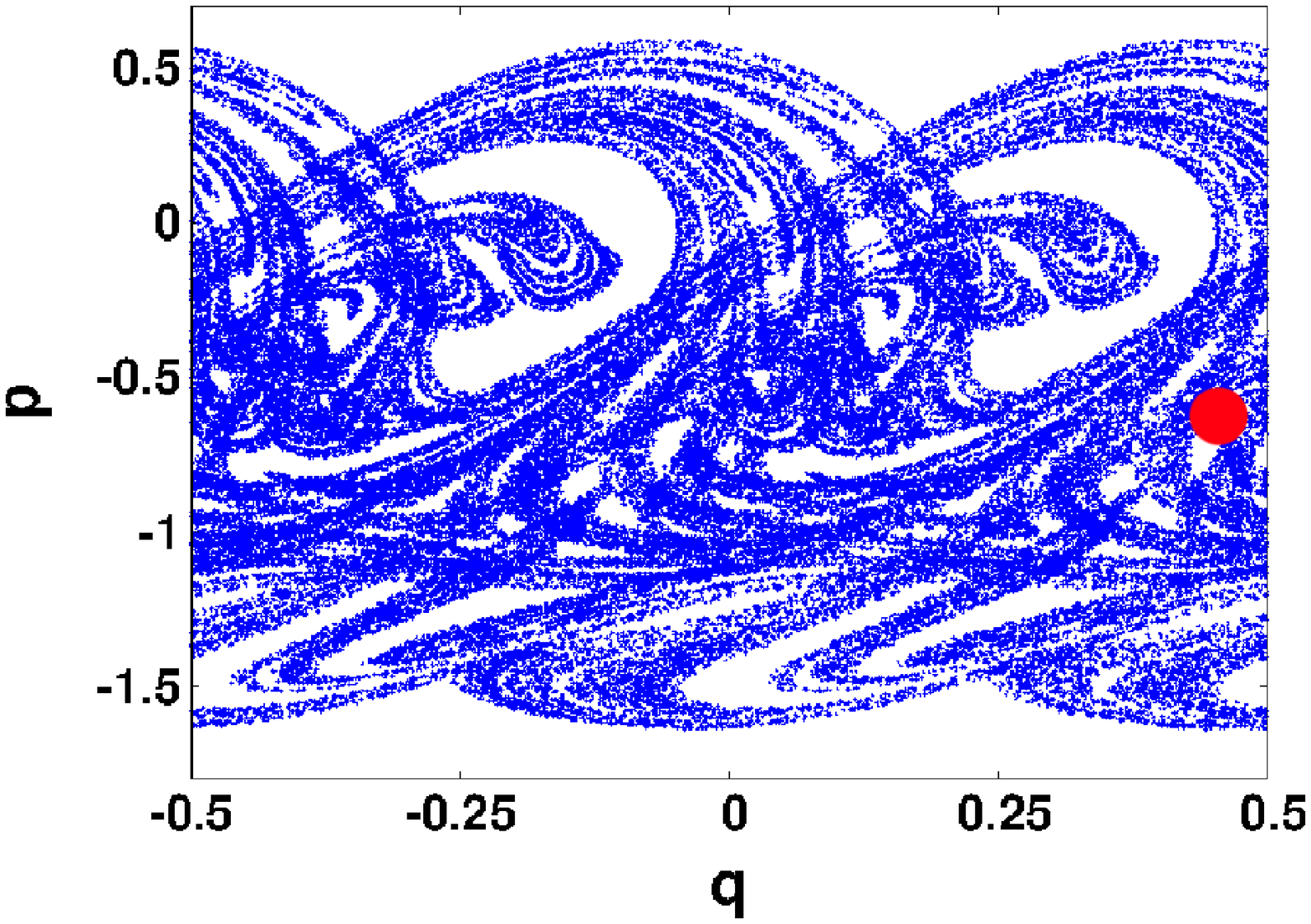}
\includegraphics[height=7.2cm, width=8.4cm]{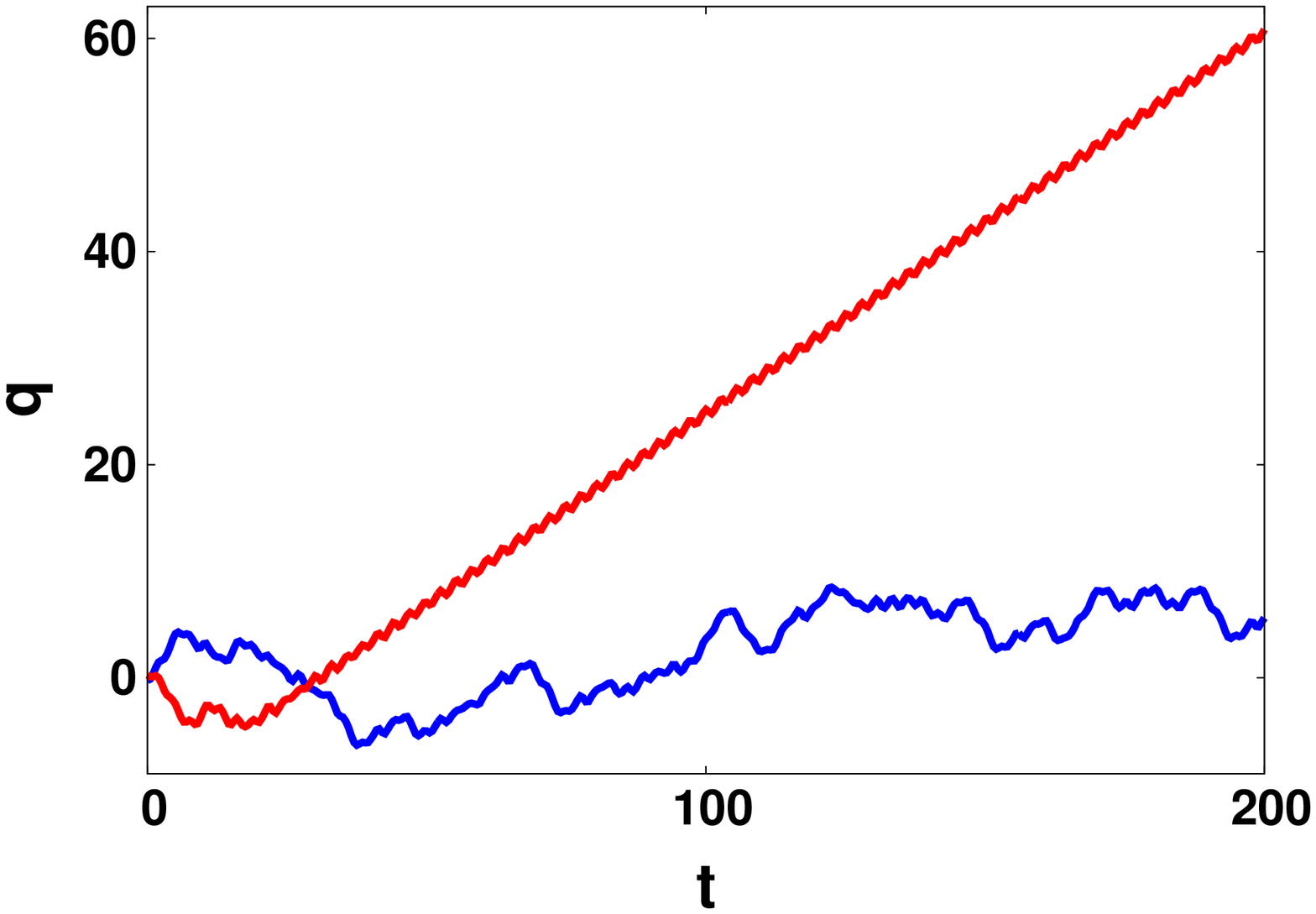}
\caption{Left panel: Stroboscopic plot showing the strange (blue) and periodic (red) attractors corresponding 
to driving strengths $h_1=1.3$ and $h_2=1.5$ respectively, in the uncoupled regime. 
The remaining system parameters are $\gamma=0.1$, $\Omega=2.25$, $\theta_0 = 0$, and $\kappa = 0.0$. 
. The dot (red) has been enlarged for emphasis. 
Right panel: A corresponding example trajectory 
for motion on the strange (blue line) and periodic (red line) attractor. Note, in the left panel the
coordinates $q$ are given $\mod(1)$.
} \label{fig:attractors}
\end{figure*}

Before exploring the system's coupled dynamics, it is important to consider the individual particle dynamics, i.e.,
the case where $\kappa =0$. For this study, $h_1$ is chosen such that the corresponding particle dynamics are chaotic,
while the choice of $h_2$ will allow for regular dynamics. 
To this end, we employ a Poincar\'e map, taking as stroboscopic time
the period of the external driving $T=2\pi/\Omega$.
The results are contained in Fig.~\ref{fig:attractors}, where the parameters used are
$\gamma=0.1$, $h_1=1.3$, $h_2=1.5$, $\Omega=2.25$, and $\kappa = 0$. Two attractors can be seen in the ($p=\dot{q}$, $q$)--plane.
The strange attractor corresponds to a driving strength of $h_1=1.3$,
while the periodic attractor results from a driving strength of $h_2=1.5$. In addition, the figure contains an example trajectory illustrating the systems uncoupled dynamics. The two dynamical regimes are clearly distinguishable.

\section{Emergence of a current}\label{sect:current}
The true complexity of this system is revealed only when the individual units making up the dimer are coupled
together, i.e. when $\kappa\ne0$. To gain a quantitative perspective on how $\kappa$ influences the
system dynamics we compute the current $\emph{v}$. More precisely, we compute the time averaged
 mean velocity for an ensemble of initial conditions, which is given by

\begin{equation}
\emph{v}= \frac{1}{T_s}\,\int_0^{T_s} dt^{\prime} \bigg\langle
\,\sum_{i=1}^{2}\dot{q}_{i}(t^{\prime})\,\bigg\rangle \,,
\end{equation}
with simulation time $T_s$ and 
with $\langle \cdot \rangle$ denoting the ensemble average. These initial conditions are chosen such that in
the uncoupled case one particle lies on the chaotic attractor, and the other on the periodic attractor seen in
Fig.~\ref{fig:attractors}. In more detail, the initial conditions corresponding to the subsystem with a
driving force of strength $h_1$ have positions $q_{1}(0)$ that are distributed uniformly over the (spatial)
period of the potential and randomly chosen velocities $\dot{q}_{1}(0)$ with $| \dot{q}_{1}(0)| \le 0.5$. The
initial conditions corresponding to the subsystem with driving strength $h_2$ were chosen so that the particle
undergoes regular rotational motion. That is, they were chosen from the basin of attraction of the forward
evolving period attractor.

\begin{figure}
\centering
\begin{overpic}[height=9.5cm, width=8.7cm]{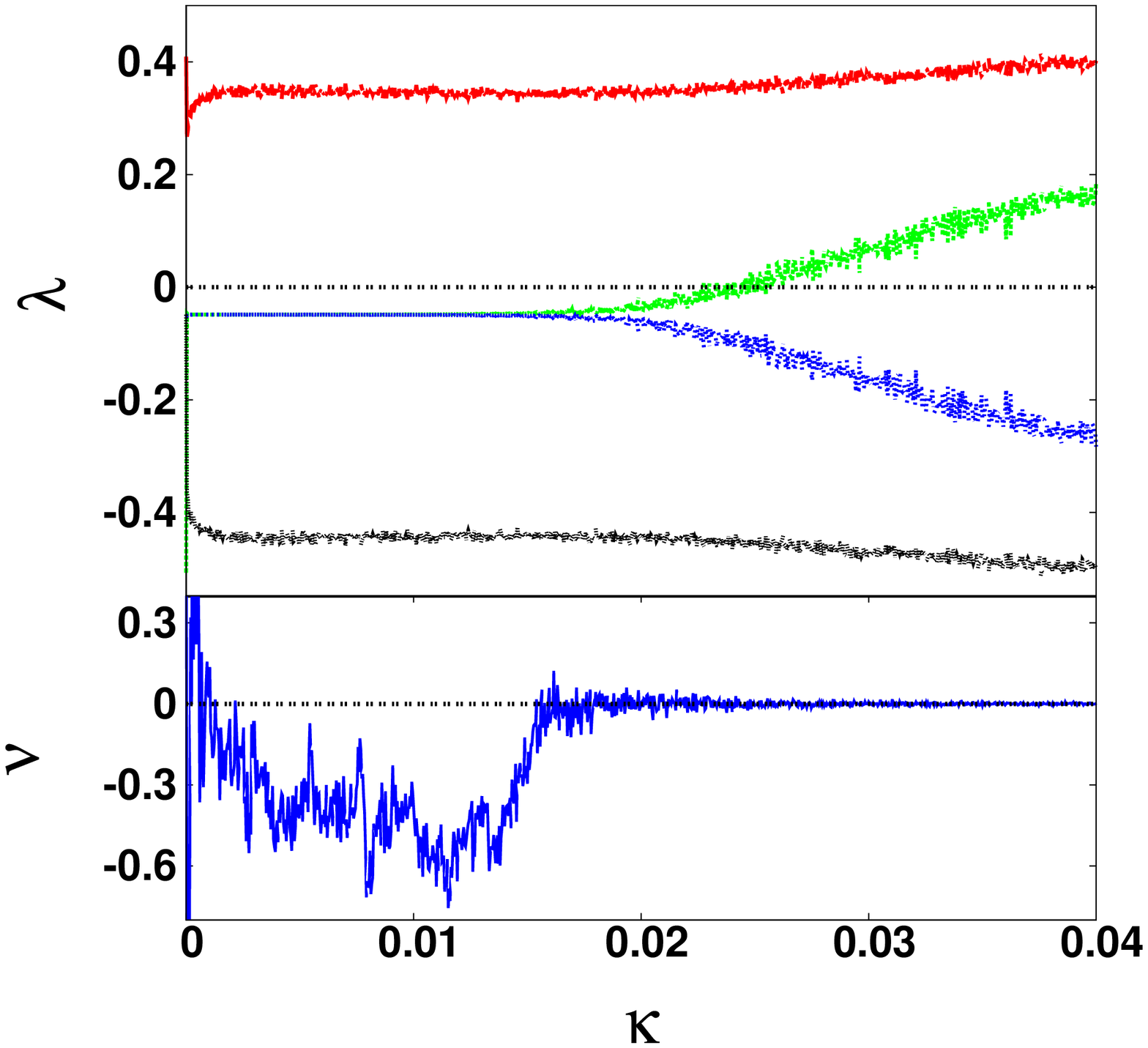}
\put(54,11){\includegraphics[height=2.0cm, width=3.0cm]{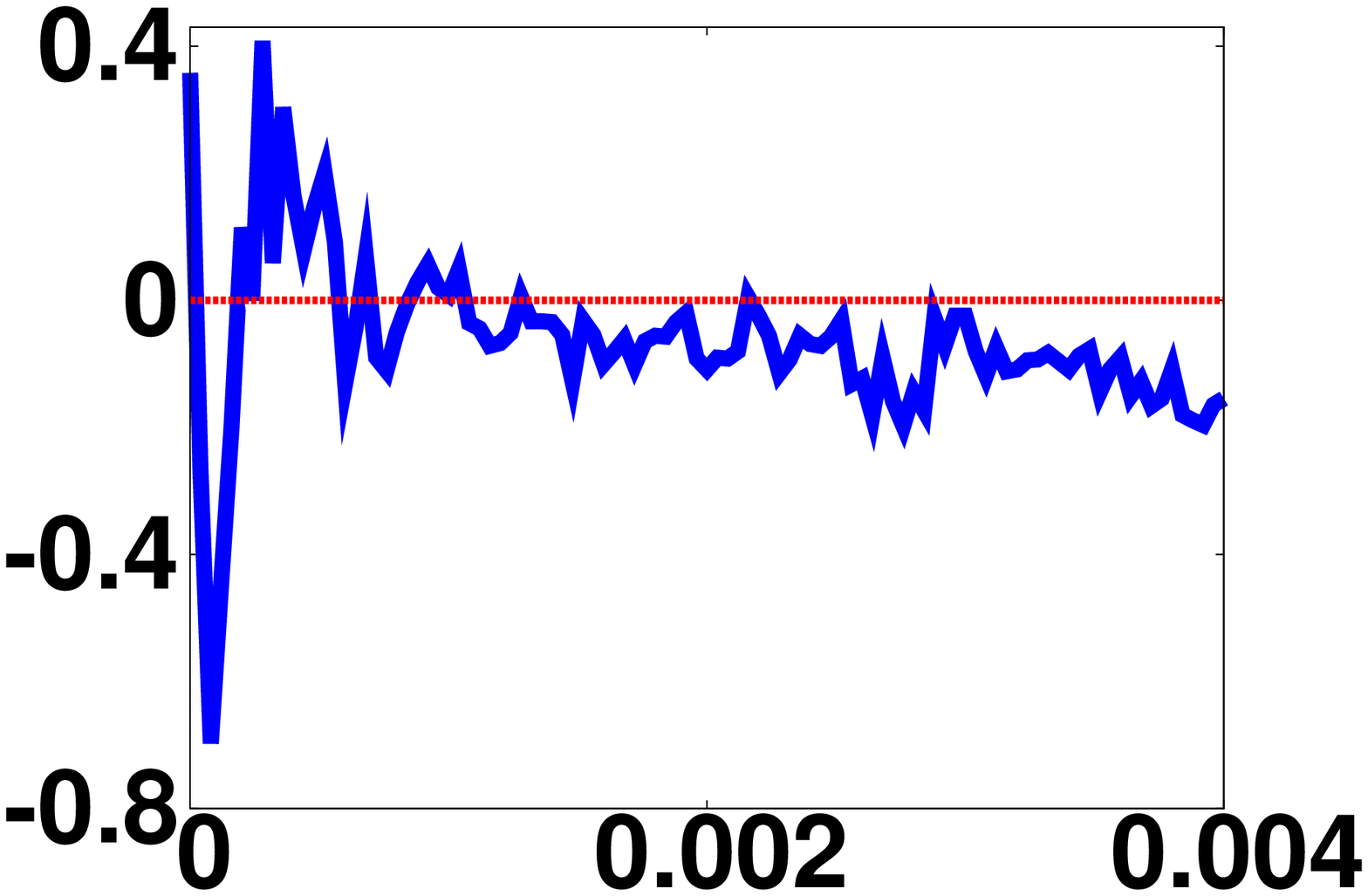}}
\put(20,71.5){\includegraphics[height=1.6cm, width=2.3cm]{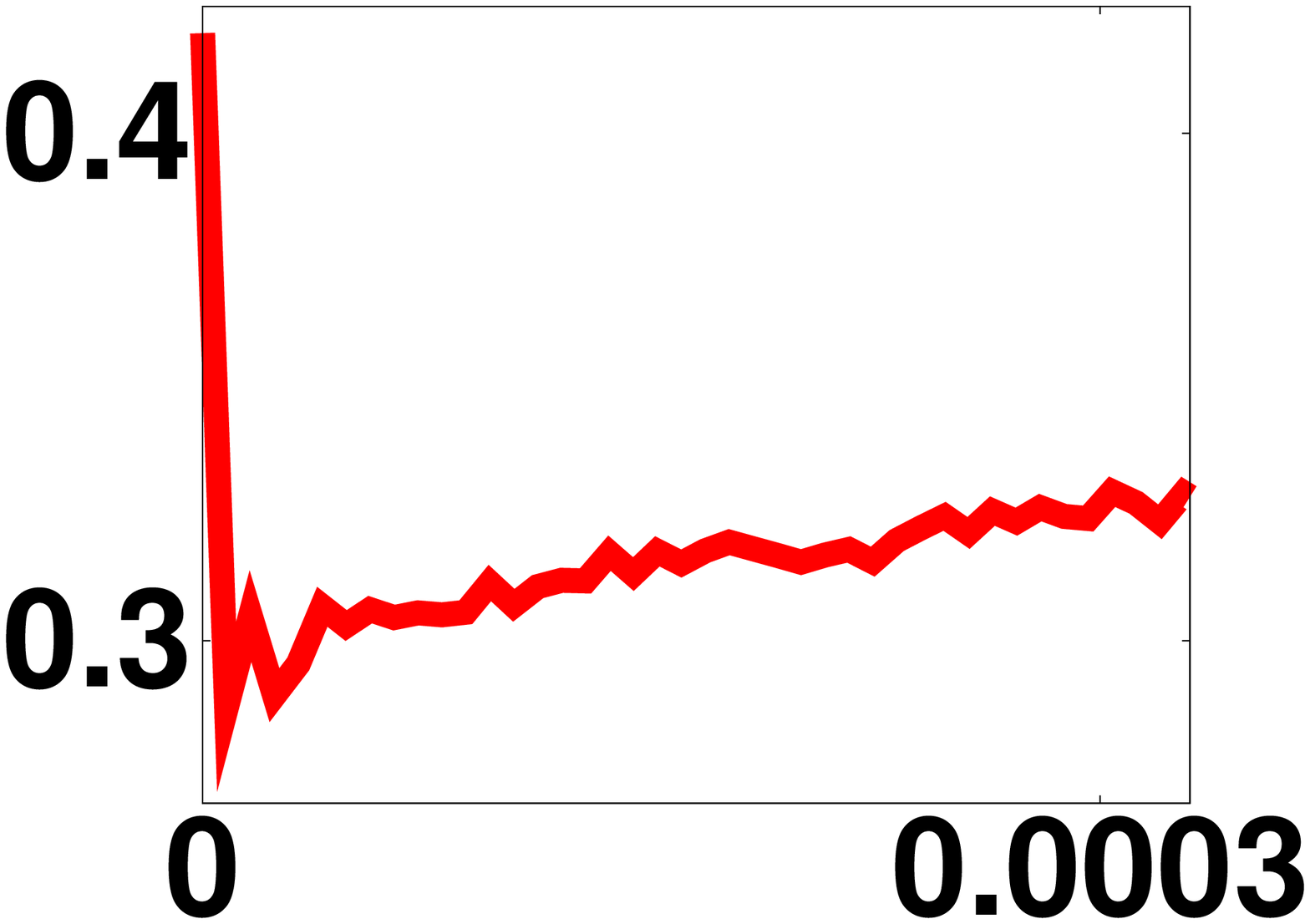}}
\end{overpic}
\caption{The particle current $\nu$, defined in section \ref{sect:current}, and the Lyapunov exponents
$\lambda$, as a function of the coupling strength $\kappa$.
The black (dashed) lines serve to guide the eye. The remaining system parameters are given in
Fig.~\ref{fig:attractors}.
The top and bottom insets show the largest Lyapunov exponent and the current, respectively, for low values of
$\kappa$.} \label{fig:current}
\end{figure}

It is clear that when $\kappa=0$ there will be a directed current supplied by the particle
on the periodic attractor. The particle on the chaotic attractor will (if at all), on long time scales,
provide
a vanishingly small contribution to the current. 
For the computation of the current in chaotic systems we refer to \cite{Kenfack}.

The key question then is, what happens to the current when the dynamics is in the coupled regime?

For computation of the long-time average, using an ensemble of $N=5000$ initial conditions, the simulation
time
interval for each trajectory is taken as $T_s=10^5$. 
(The simulation time $T_s$ exceeds, by magnitudes of order, the typical time scale of the system 
given by the period duration of oscillations near the bottom of a potential well amounting to 
$T_0=2\pi/\omega_0=\sqrt{2\pi}$.)
Further, the current was calculated for $1000$ (evenly spaced) values of $\kappa$ 
in the range $0 \le \kappa \le 0.04$.
The results are contained in the bottom panel of Fig.~\ref{fig:current}.

As expected, for $\kappa=0$ there is indeed a directed current. With low values of $\kappa$ a number
of current reversals can be observed (shown on bottom inset in Fig.~\ref{fig:current}). In the window $0.001
\lesssim \kappa \lesssim 0.015$ a negative current is supported, 
followed by a transition to zero. It is this transition, and the subsequent negligible current which is of
most interest to us. To explain this we will look at the Lyapunov spectrum for the system given by the
equations of motion Eqs.~(\ref{eq:q1})-(\ref{eq:q2}).

The Lyapunov spectrum was computed using the method of Gram-Schmidt orthonormalisation \cite{parker}, which
allows for the simultaneous estimation of all of the systems Lyapunov exponents. The results are contained in
the top panel of Fig.~\ref{fig:current}. As was mentioned earlier, the passing from an uncoupled to a coupled
regime can be seen to mark the beginning of a competition between two dynamical regimes. With very weak
coupling it is the regular dynamics that prevail. Not only can a directed current be found, but also the
effect of chaos is, to an extent, suppressed. Looking more closely at the only positive Lyapunov exponent 
in this regime (inset Fig.~\ref{fig:current}), we see that when the particles become coupled an instant
reduction in the magnitude of the exponent follows. For low values of $\kappa$ the spectrum remains almost
constant with one positive and three negative Lyapunov exponents. In the range $0.016 \lesssim \kappa \lesssim
0.024$ two of the exponents diverge from one another. Crucially, this divergence 
sees one of the exponents approach zero. This divergence also coincides with a reduction in the magnitude of
the current (see Fig. \ref{fig:current}). At a critical coupling value $\kappa_c \approx 0.024$ a second
exponent becomes positive marking the transition to hyperchaos. Looking again at Fig. \ref{fig:current}, it
can be seen that the current is almost negligible. Thus, the transition to hyperchaos has removed any inherent
bias in the system and now no direction of motion is favoured, i.e. the possibility of a directed current is
excluded. In the next section we will discuss phase-space structures and how they change with increasing
$\kappa$, and in addition relate them to the current and Lyapunov spectrum discussed in this section.

\begin{figure}
\centering
\begin{overpic}[height=9.3cm, width=8.7cm]{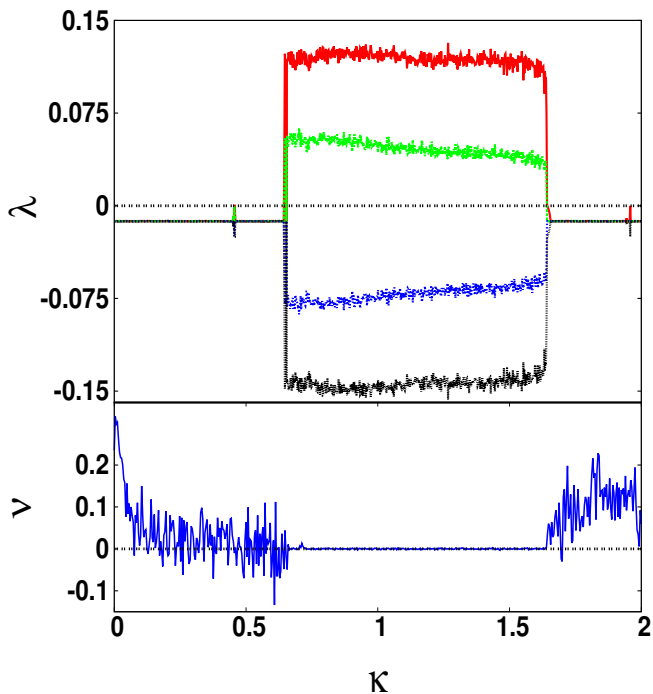}
\end{overpic}
\caption{The particle current $\nu$, defined in section \ref{sect:current}, and the Lyapunov exponents
$\lambda$, as a function of the coupling strength $\kappa$.
The black (dashed) lines serve to guide the eye. The system parameters are the same as given in
Fig.~\ref{fig:attractors}, except for $h_1=h_2=1.5$.} \label{fig:current2}
\end{figure}

Before moving on, it is worth discussing the generality of the results presented in this section. First of
all, scanning the entire parameter space is (if possible at all) practically infeasible. Moreover, a change in
parameters has implications for the location and types of attractors that can be found in phase-space.
However, simulations reveal that the hyperchaos - zero current relation persists when the parameters are
chosen from the regular and periodic windows observed in the bifurcation diagram of the single particle. For
illustrastion, Fig.~\ref{fig:current2} shows the current and Lyapunov spectrum, as a function of the coupling
strength, for a system that is qualitatively different from that which is the focus of this paper. In more
detail, this system has two subsystems were the underlying attractors are \emph{both} regular. Importantly,
the regime of hyperchaos coincides with the window of zero current.

\section{phase-space Structures}\label{sect:pss}
We have seen in the uncoupled regime, i.e. non-interacting particles, that it suffices to examine
(stroboscopically) the
phase-space for each particle individually (see Fig. \ref{fig:attractors}). 
However, when the particles are coupled the phase-space becomes five dimensional and such representations,
yielding four dimensional Poincar\'{e}
surfaces of section (PSSs), are no
longer suitable. We instead will present three dimensional projections of the four dimensional PSSs
illustrating the changes in phase-space through
increased coupling. Further, for each $\kappa$ there are four possible three dimensional projections that
illustrate qualitatively
the same point, and thus we will omit three of them.  

\begin{figure*}[ht!]
\includegraphics[height=4.5cm, width=5.8cm]{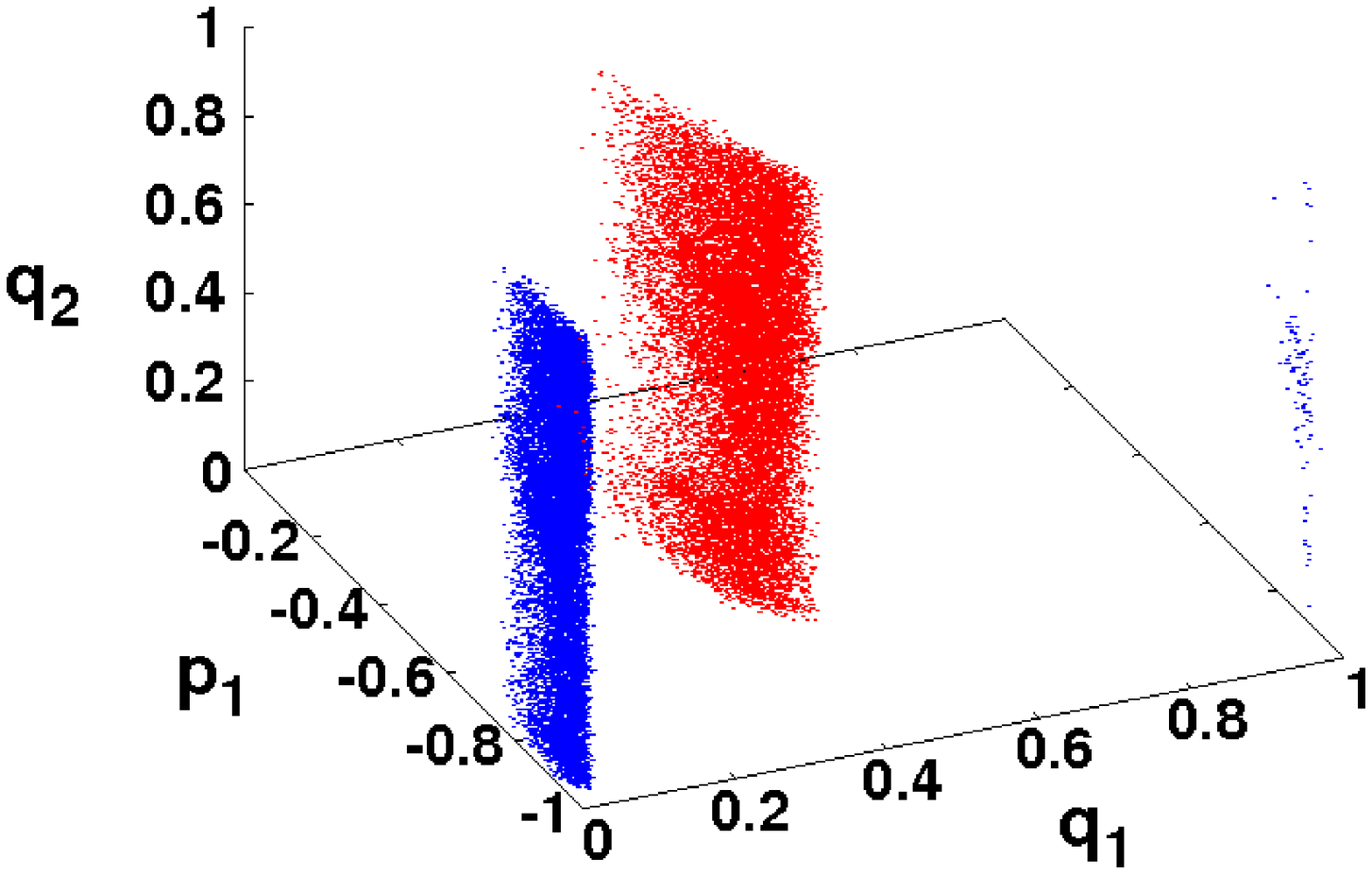}
\includegraphics[height=4.5cm, width=5.8cm]{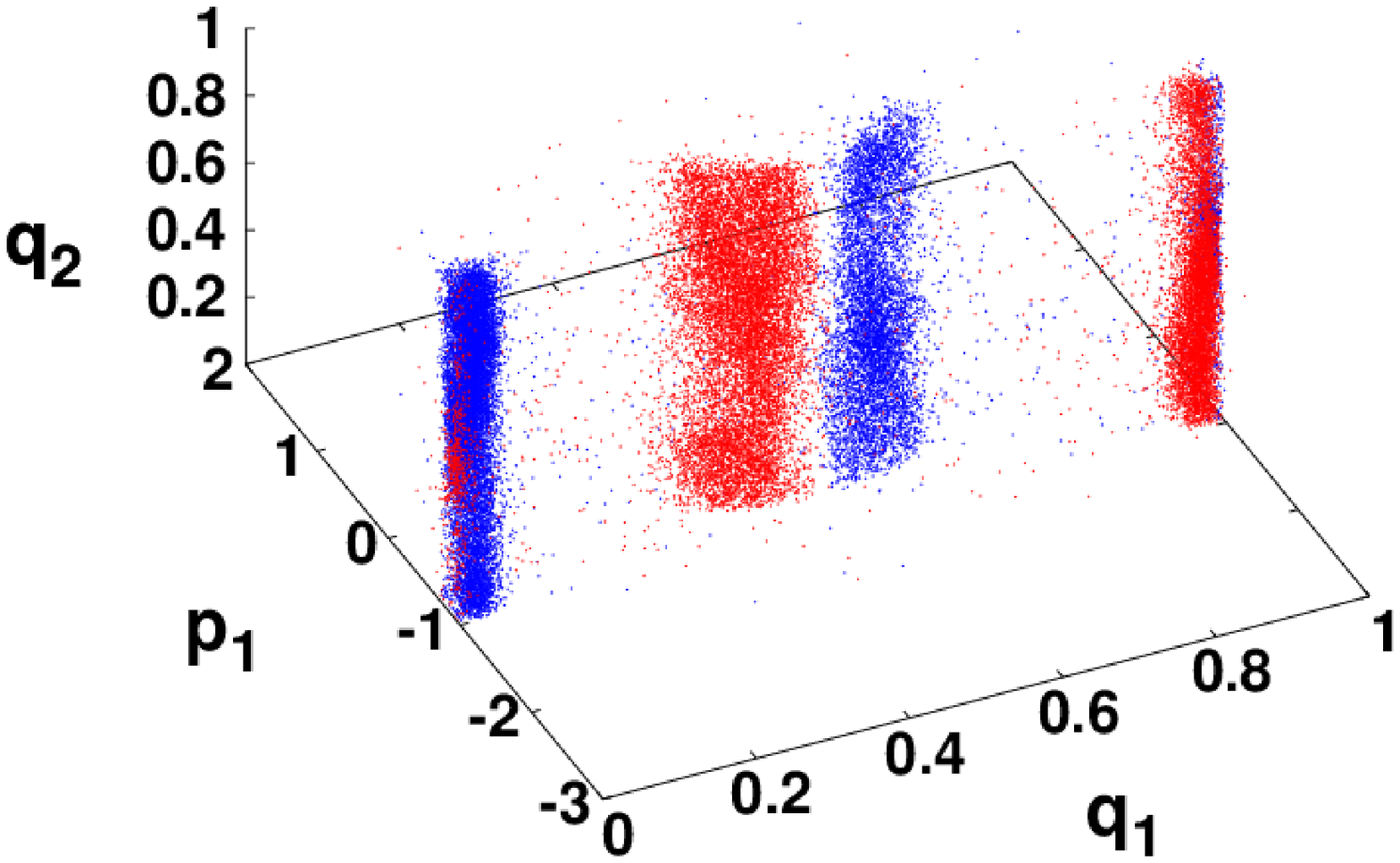}
\includegraphics[height=4.5cm, width=5.8cm]{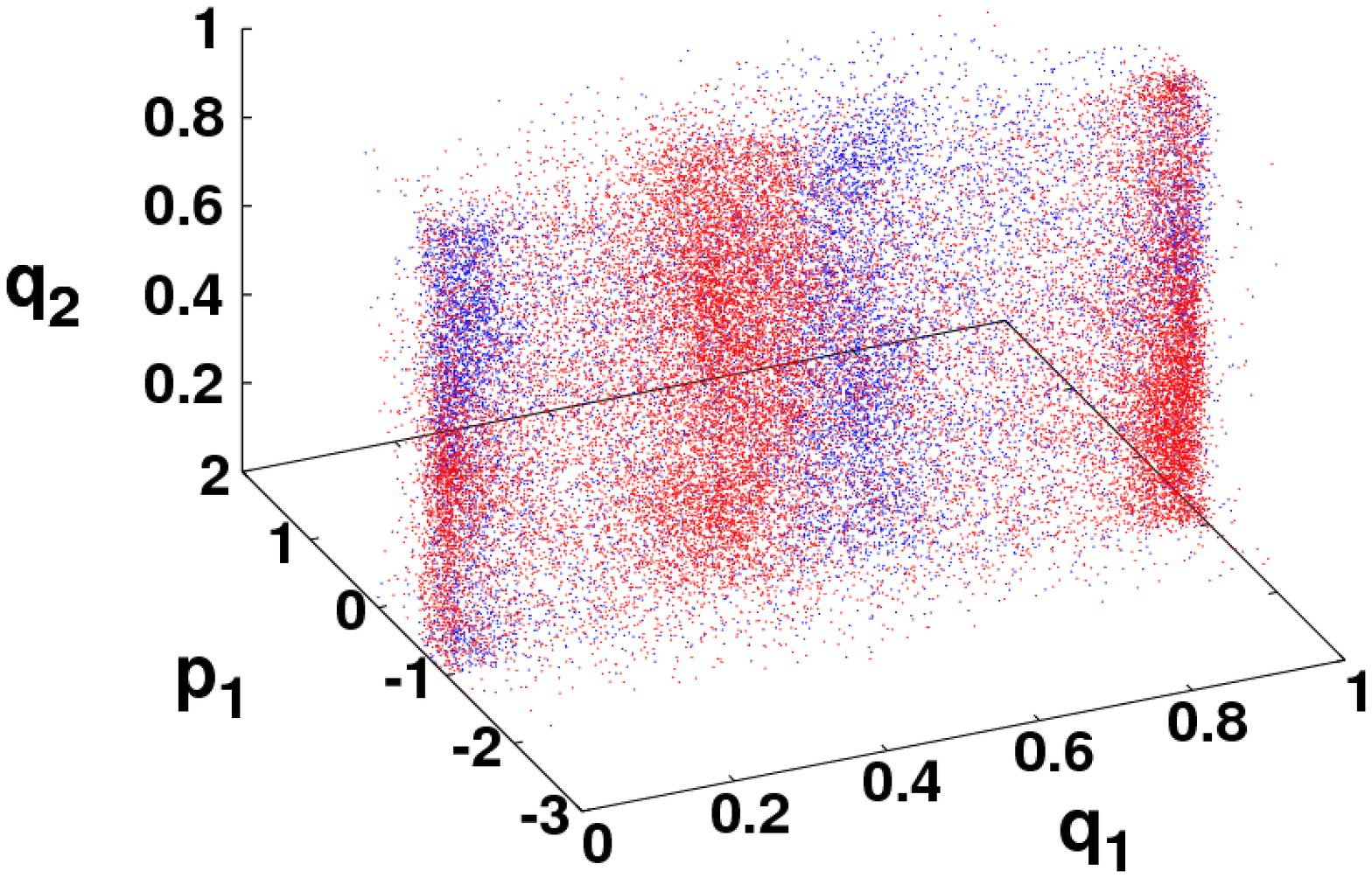}\\
\begin{overpic}[height=4.5cm, width=5.8cm]{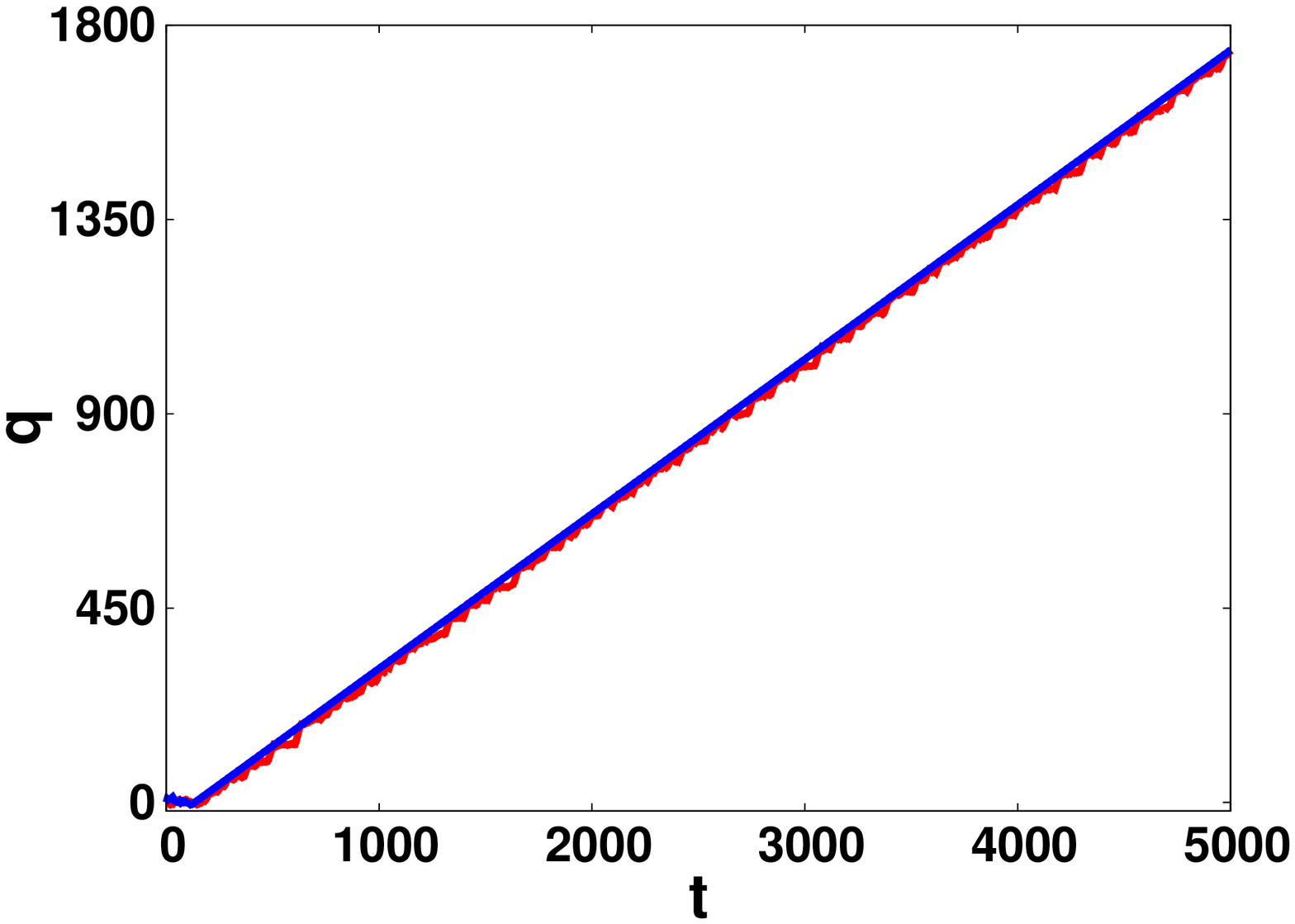}
\put(46.5,11){\includegraphics[height=1.8cm, width=2.7cm]{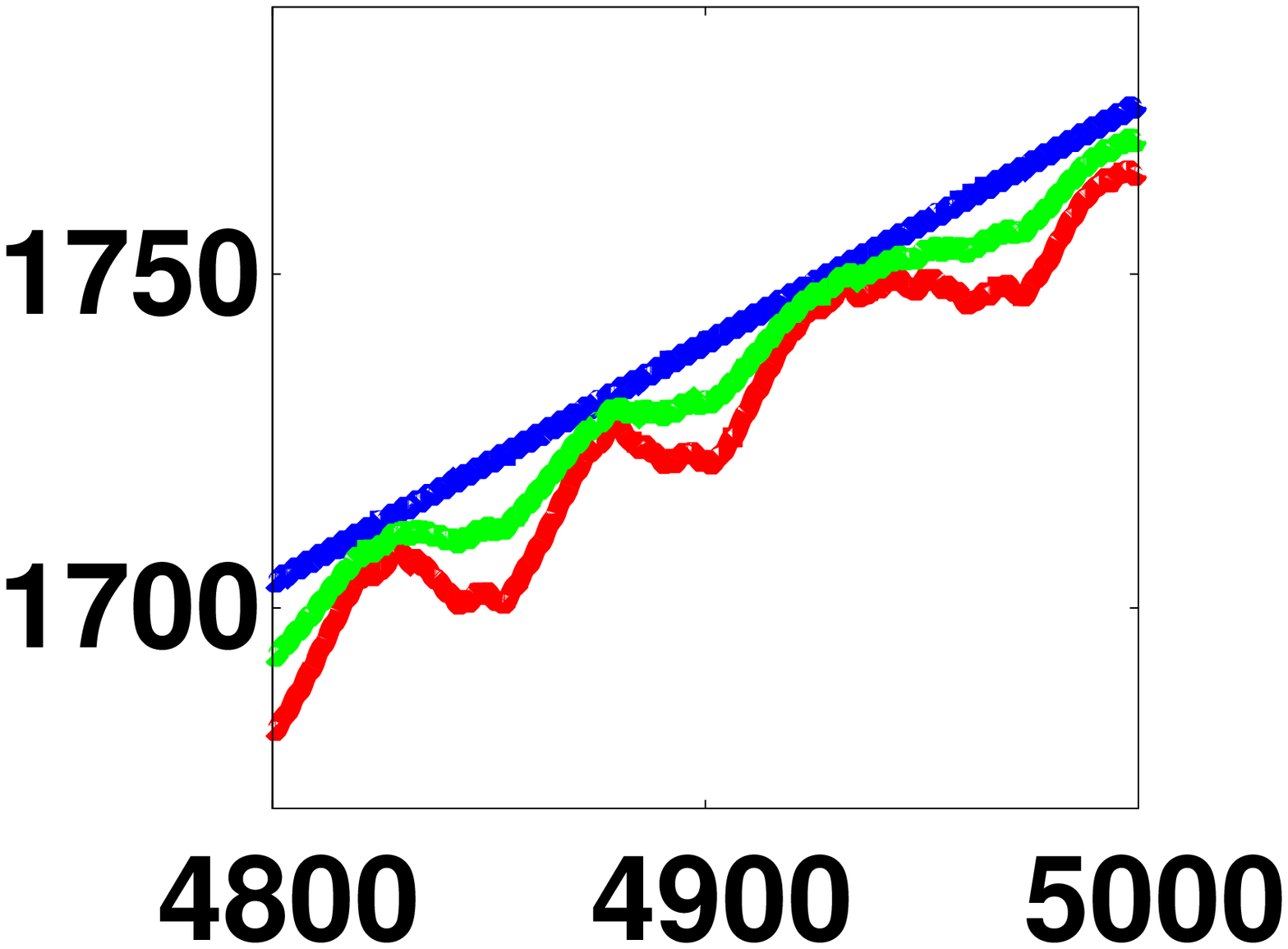}}
\end{overpic}
\begin{overpic}[height=4.5cm, width=5.8cm]{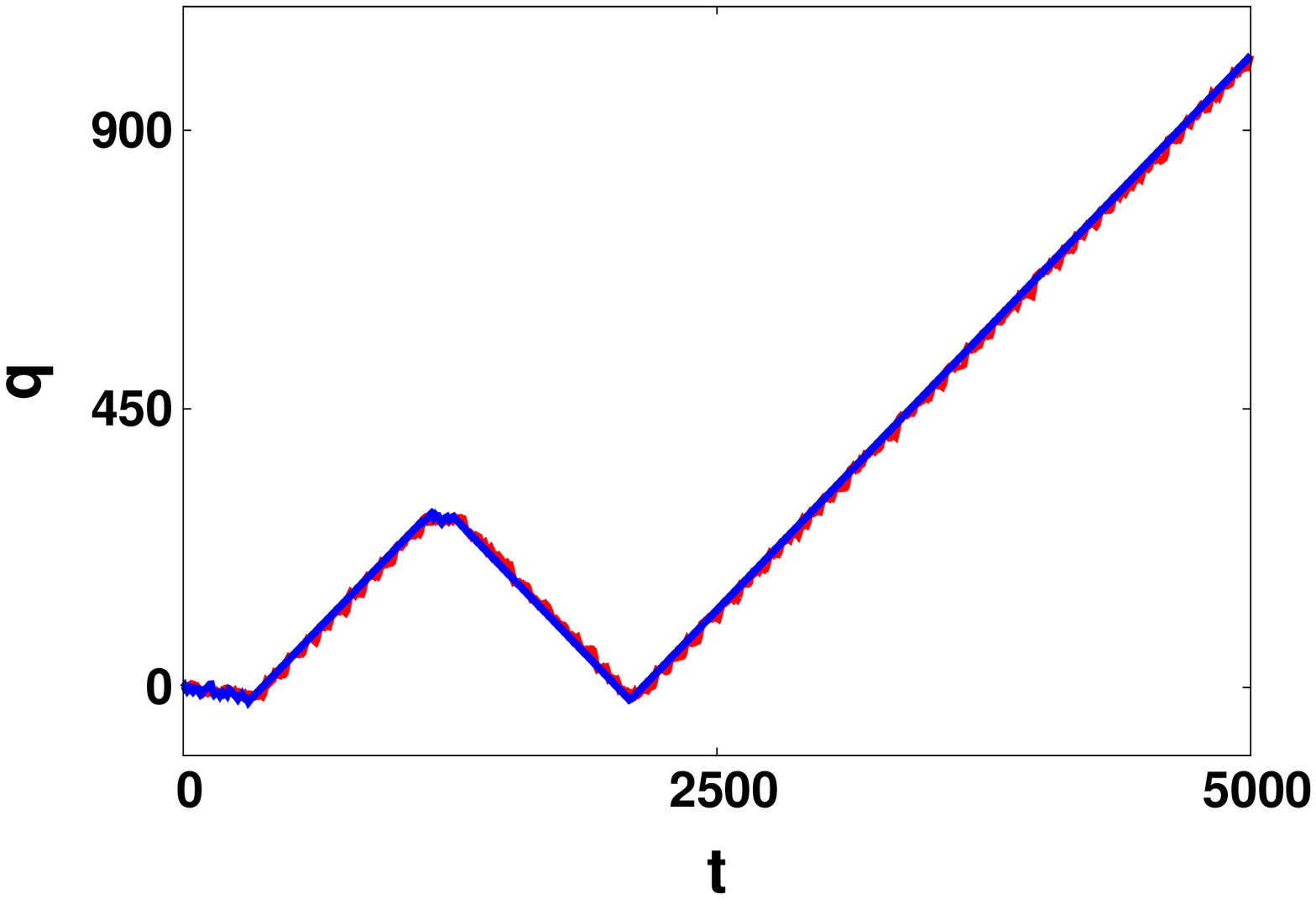}
\put(27,42){\includegraphics[height=1.9cm, width=2.6cm]{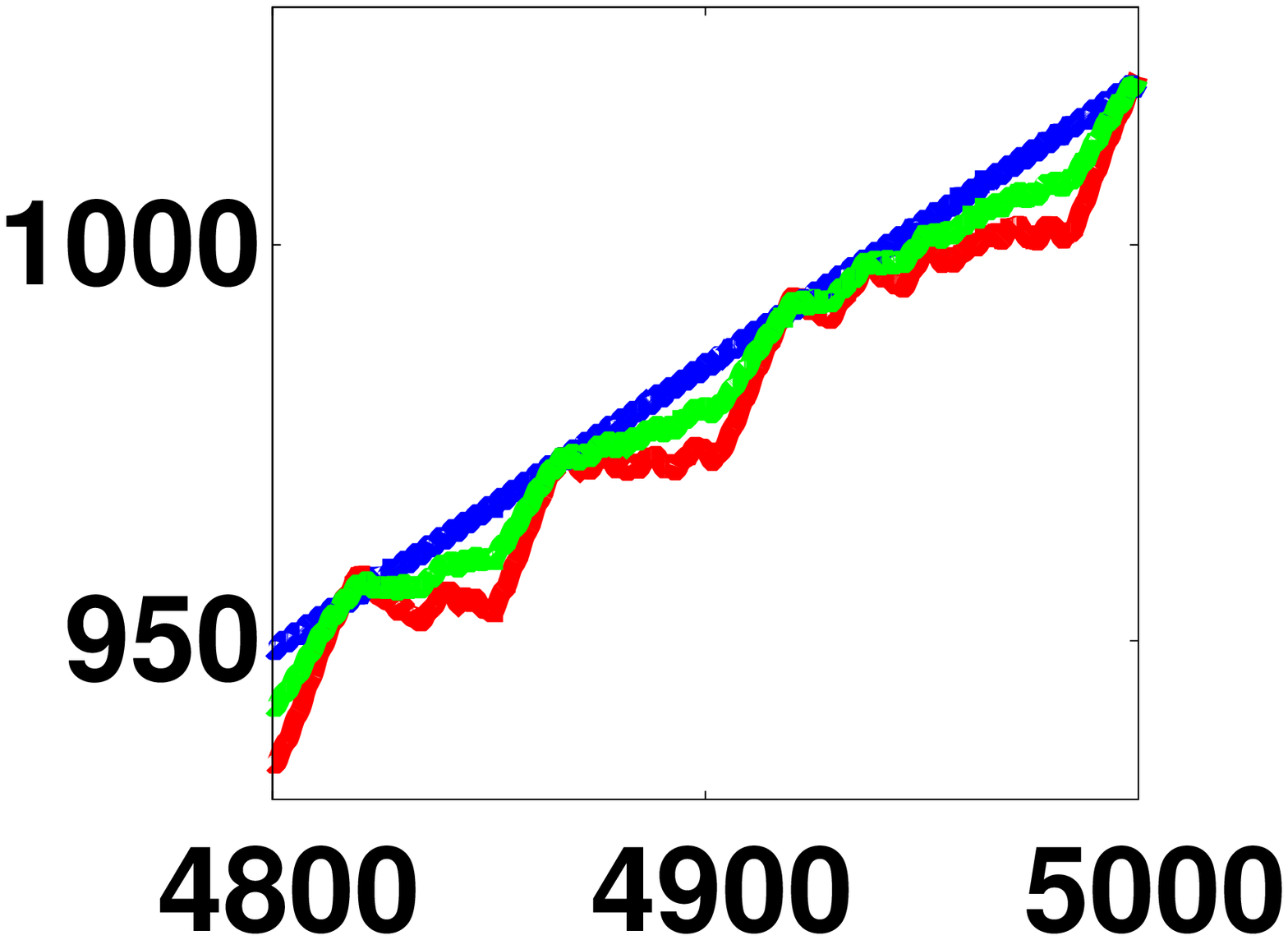}}
\end{overpic}
\begin{overpic}[height=4.5cm, width=5.8cm]{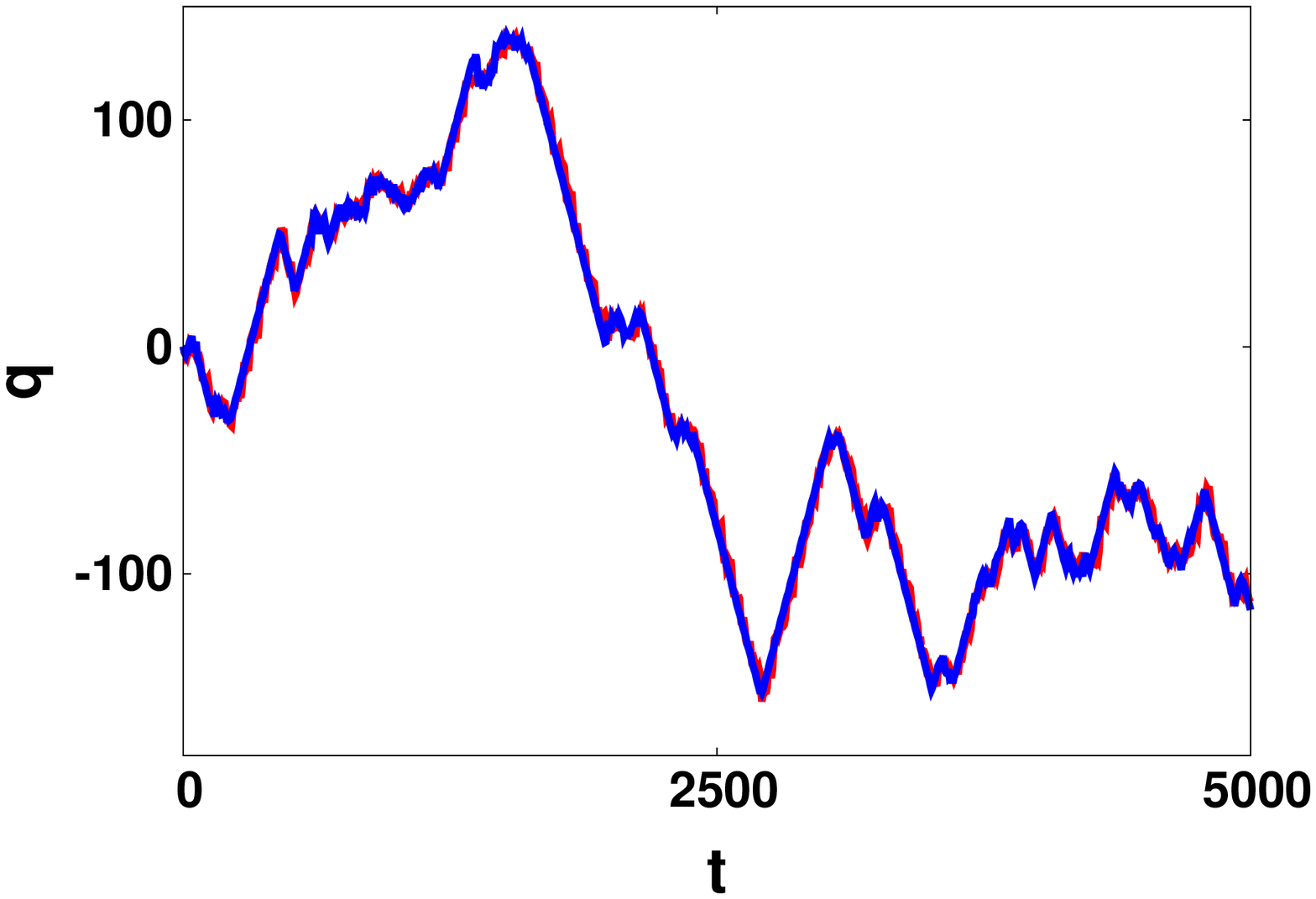}
\put(50,49){\includegraphics[height=1.5cm, width=2.5cm]{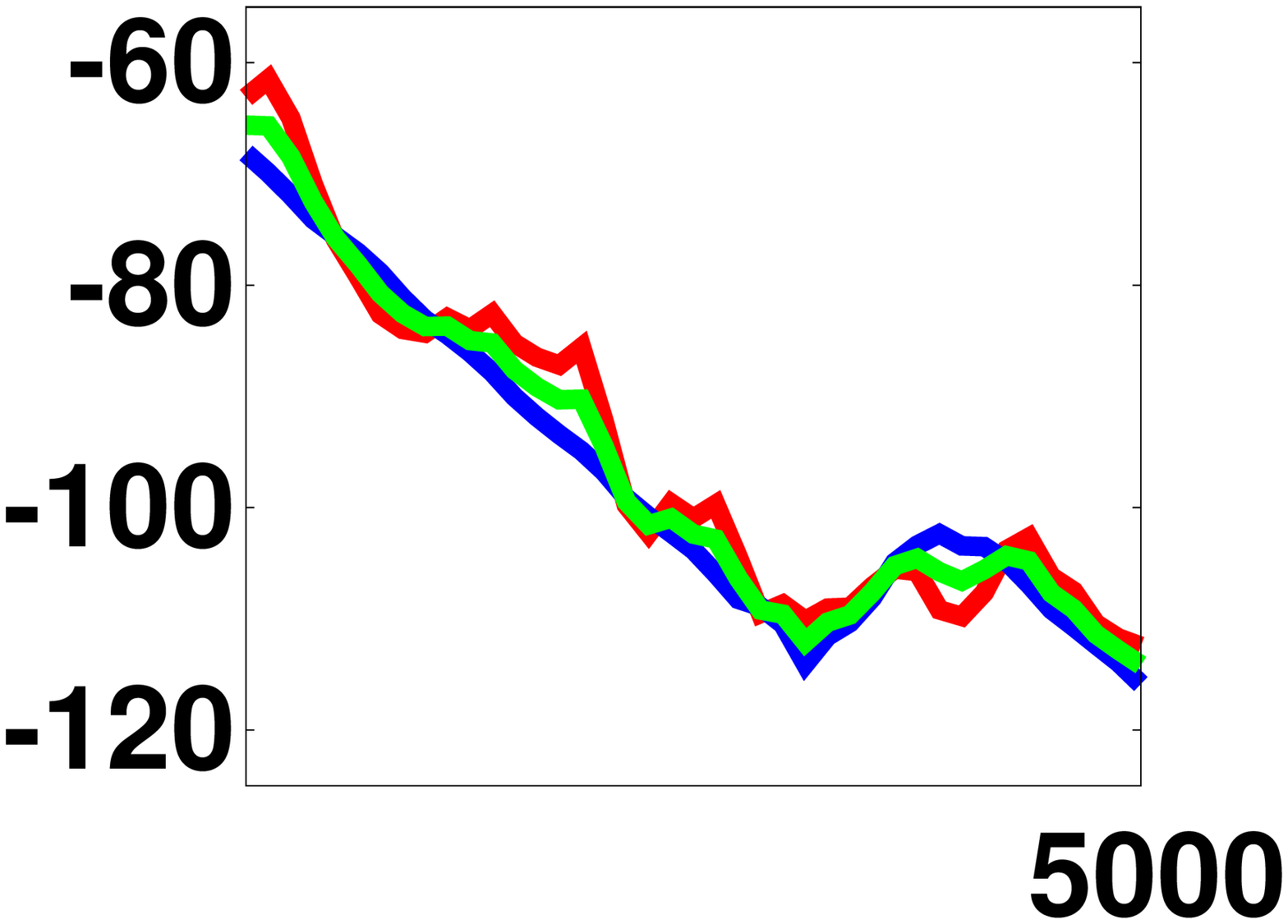}}
\end{overpic}\\
(a) \hspace{5.5cm}(b) \hspace{5.5cm}(c)
\caption{Top row: Three dimensional projections of the attractors in phase-space for different coupling
strengths. 
Left: $\kappa = 0.01$, Middle: $\kappa = 0.02$, Right: $\kappa = 0.04$. In the left panel the attractor shown
in blue (red) provides 
coherent transport to the right (left). In the middle and right panel blue (red) points belong to such orbits
whose initial conditions
are situated in the basin of attraction of a transport-providing attractor shown in the left panel in blue
(red).
The remaining parameter values are
given in Fig. \ref{fig:attractors}. Bottom row: Examples of corresponding trajectories with the centre of mass
shown in green.} \label{fig:pss3d}
\end{figure*}

Fig.~\ref{fig:pss3d}(a) shows one of the three dimensional projections for the case of weak coupling,
$\kappa=0.01$. It shows
two distinct attractors (each having its own basin of attraction which are separated by impermeable basin
boundaries) that take the particles in opposite directions. 
Strikingly, although these are chaotic attractors, indicated by 
a positive Lyapunov exponent (cf. previous section), the motion on them is nevertheless effectively directed.
Like in  the uncoupled case, for the system driven at $h_2=1.5$, the different weights attached to these
transporting 
attractors favour a directed
current (Fig. \ref{fig:current}), a feature that is maintained up to coupling strengths $\kappa \lesssim
0.016$.
We recall that in a single particle system, i.e. 
$\kappa=0$, 
the  directed motion results from a
lowering of the dynamical symmetry caused by the external
modulation field \cite{Yevtushenko},\cite{single}. That is, even though the
potential and the external driving field are symmetric with respect to space and time respectively, with the
choice of a  fixed
phase $\theta_0$ the symmetry of the flow is reduced and a
phase-dependent net motion is found. Due to symmetry reasons it holds that the sign of the mean velocity 
 is reversed upon the changes $\theta_0=0\,\rightarrow
\theta_0=\pi$ and $h_2\rightarrow -h_2$ respectively. However, there
exists a phase $0<\theta_0<\pi$ for which symmetry between the two
coexisting periodic attractors, supporting solutions with
velocities of opposite sign, $v_2$ and $-v_2$, is restored and
therefore the net motion vanishes. Hence, in this case for the coupled system the two attractors and their 
respective basins of attraction are symmetric in phase-space.

As previously mentioned, the interesting feature of Fig.~\ref{fig:current} was the transition
to, and subsequent negligible current (with increasing $\kappa$). When the coupling strength, playing the role
of the bifurcation parameter, is increased the three dimensional projections of the PSSs reveal that the two
attractors get larger and eventually merge  (cf. \ref{fig:pss3d}(b)) and \ref{fig:pss3d}(c)).
That is, a merging crisis takes place for which, at a critical value 
$\kappa_m\simeq 0.015$, the two enlarged attractors collide simultaneously 
with the basin boundary
which separates their basins of attraction. As a result they eventually merge, after the crisis, forming
a single large chaotic attractor in phase-space \cite{ott},\cite{Grebogi}. Note for the 
coupling strength $\kappa_m \simeq 0.015$,  when the crisis occurs, there is
 only one positive Lyapunov exponent. The beginning of the crisis coincides with the event when two of the
Lyapunov exponents diverge from 
 each other, with one of them rapidly approaching zero upon increasing the value of $\kappa$ (corresponding to
the 
 green and blue line in Fig.~\ref{fig:current}). At the same time the current rapidly approaches zero.
It is also worth noting that if just one of the four three-dimensional projections shows
separated attractors, then the attractors are separated in the four-dimensional phase-space.
Conversely, to observe a merging of the attractors in the four-dimensional phase-space, it requires
that all four three-dimensional projections show a merging of the attractors.

\begin{figure*}[ht!]
\includegraphics[height=4.8cm, width=5.8cm]{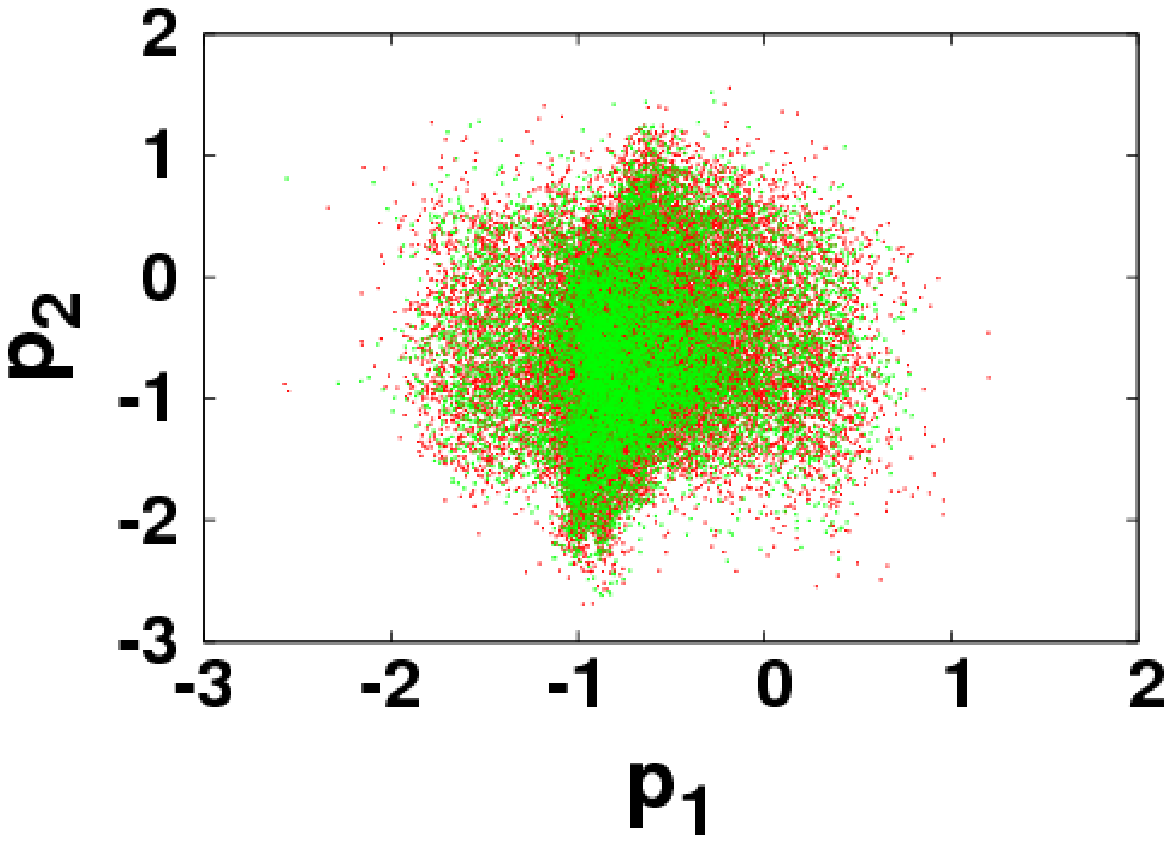}
\includegraphics[height=4.8cm, width=5.8cm]{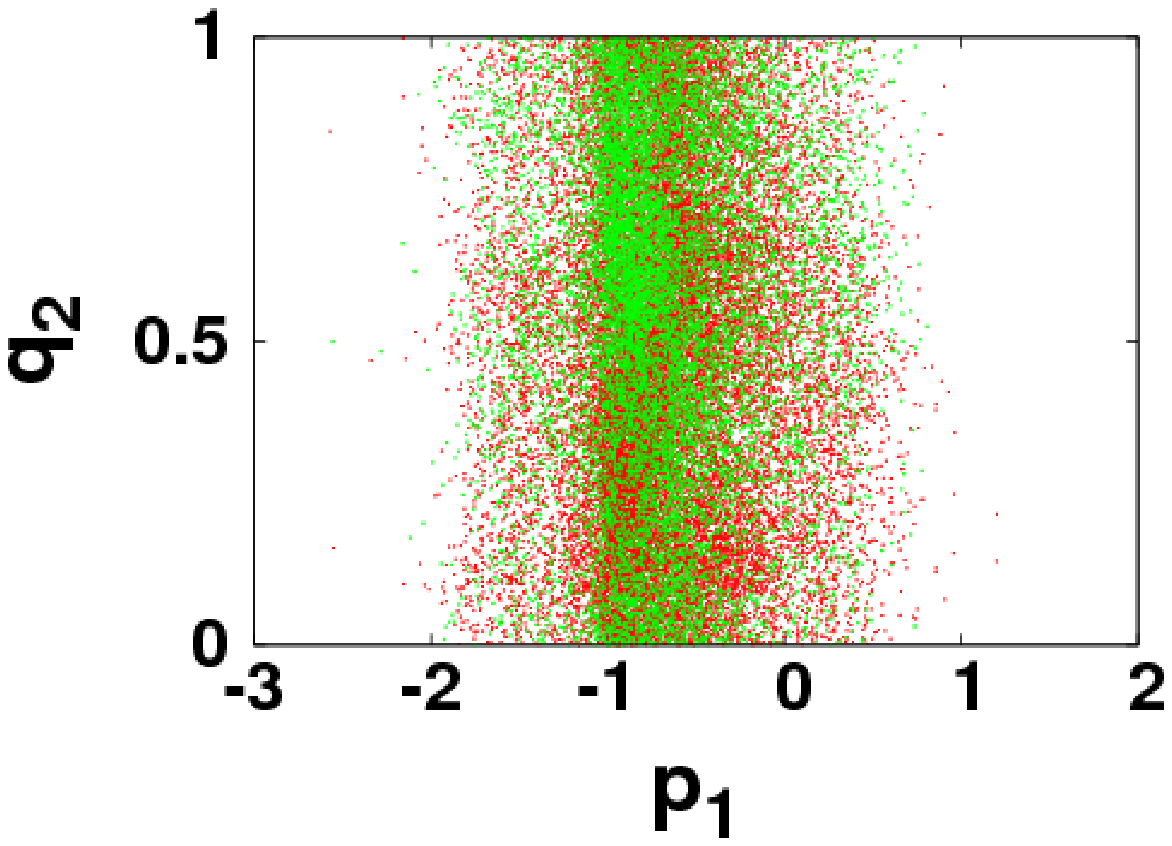}
\includegraphics[height=4.8cm, width=5.8cm]{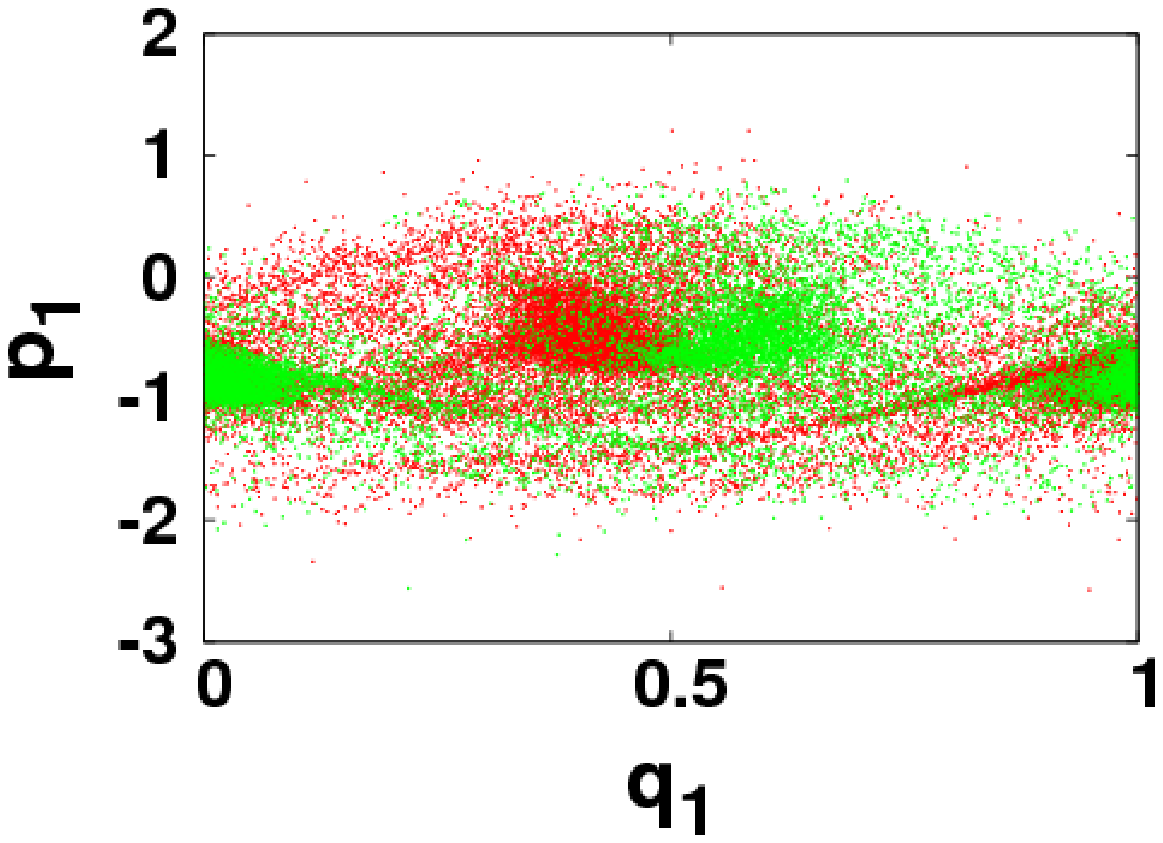}\\
\includegraphics[height=4.8cm, width=5.8cm]{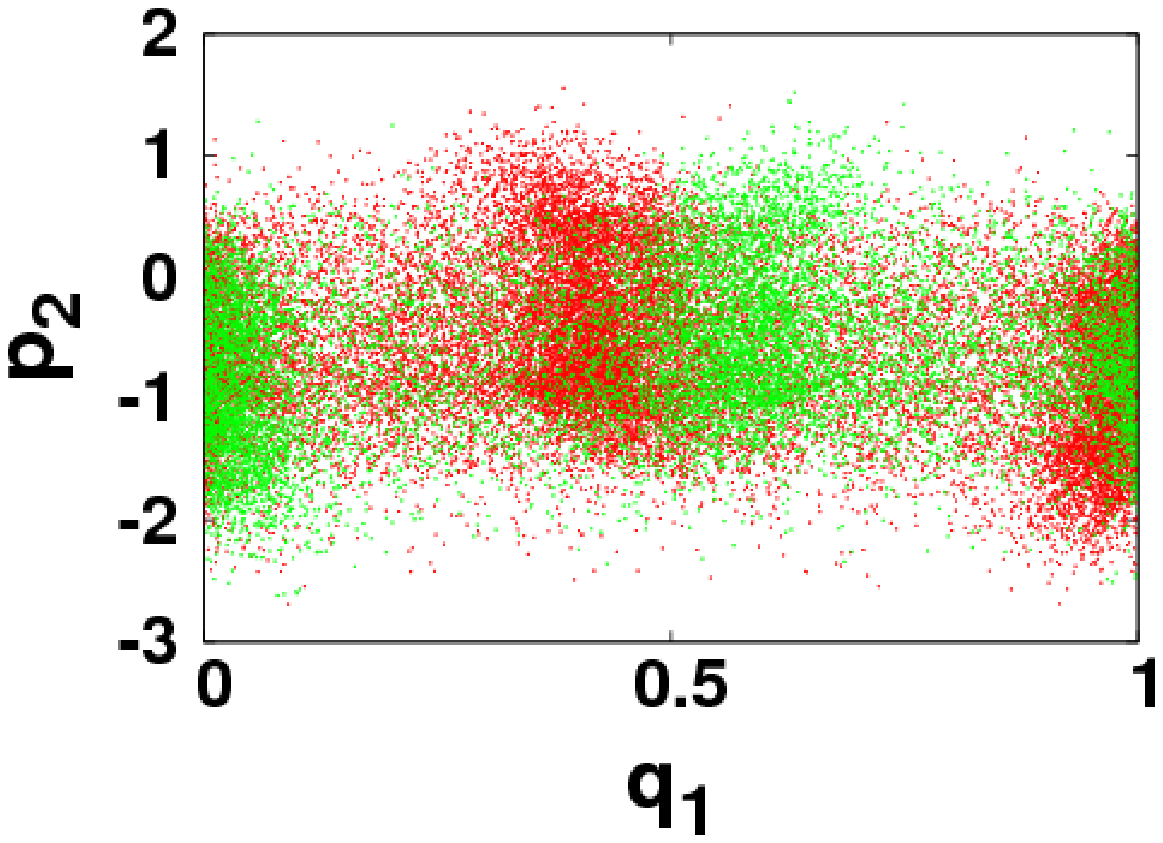}
\includegraphics[height=4.8cm, width=5.8cm]{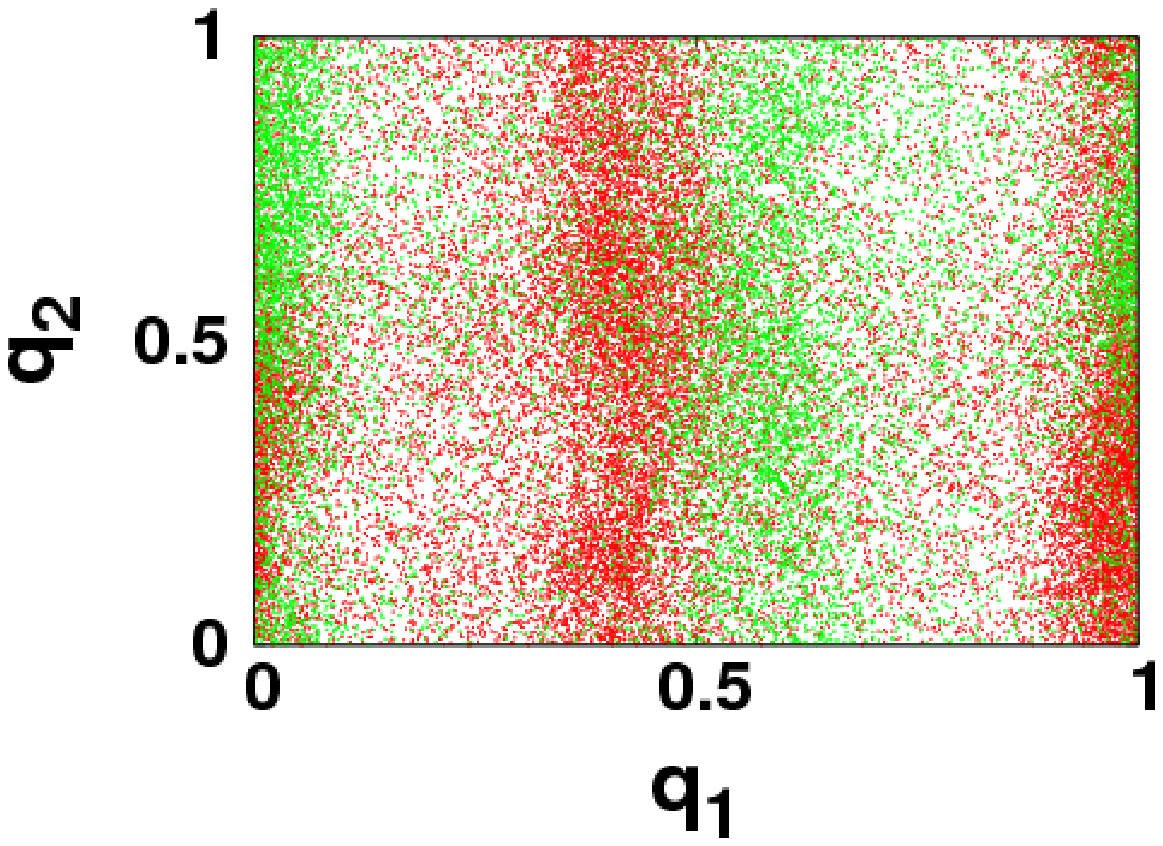}
\includegraphics[height=4.8cm, width=5.8cm]{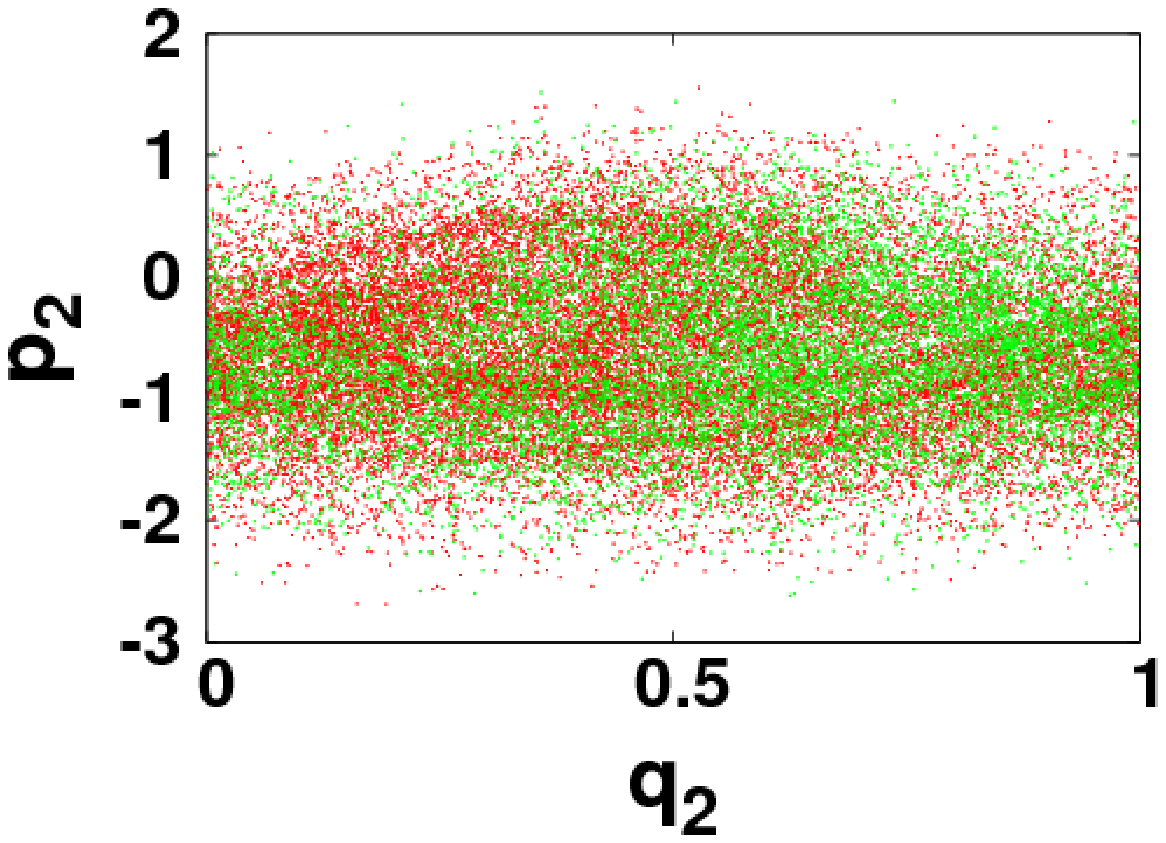}

\caption{All possible two-dimensional projections (excluding reflections) showing the merging of the two attractors for the case $\kappa=0.4$. The remaining parameter values are given in Fig. \ref{fig:attractors}.} \label{fig:pss2d}
\end{figure*}

Moreover, for under-critical values of $\kappa_m \lesssim \kappa \lesssim \kappa_c$, when the degree of
merging is not too 
pronounced (cf. Fig.~\ref{fig:pss3d}(b)) 
so that the structure of the formerly separated attractors is still discernible, trajectories jump in a random
fashion from the remnant of one 
of the attractors, after having spent a considerable amount of time there,  to the other one and vice versa.
As the coupling strength is increased these
sojourn times get shorter and shorter.  
This is illustrated in Fig. \ref{fig:pss3d}(b), 
where, as the result of growth and more pronounced merging, the attractor is  covering larger regions of
phase-space and the associated 
net current  is already very close to zero. 
The final three dimensional
projection Fig. \ref{fig:pss3d}(c) with $\kappa=0.04$ shows the phase-space when hyperchaos is well
established.
In addition, this coupling strength corresponds to a vanishingly small current. The two attractors have now
completely
merged indicating that no direction of motion is favoured. In other words, symmetry between forward and
backward motion is restored.
To demonstrate that the attractors have completely merged, we present in Fig. \ref{fig:pss2d} all possible
two-dimensional
projections of the four-dimensional phase-space.

The features of the attractors before and after the merging crisis are reflected in the time evolution of the
trajectories displayed  
in Fig.~\ref{fig:pss3d}. In more detail, for coupling strength $\kappa=0.01$ the trajectory in the left panel
of Fig.~\ref{fig:pss3d}
resembles a transporting periodic trajectory of the uncoupled regime.
Only a magnification of the plot reveals the chaotic wiggling of the trajectories around the seemingly
straight line
 corresponding to unidirectional particle transport. As the two attractors in phase-space
 are separated the motion of the particles, associated with a pair of trajectories captured by  one of the
attractors, proceeds unidirectionally.

In contrast, for coupling strength $\kappa=0.02$, i.e. after the merging crisis, the pair of trajectories
shown in Fig.~\ref{fig:pss3d}\,(b)
undergoes sudden changes in direction belonging in phase-space to 
transitions between the  `skeletons' of the two attractors having been transporting before the merging crisis.
Finally, in the hyperchaotic regime 
for $\kappa=0.04$ the pair of trajectories exhibits no coherent properties at all (cf.
Fig.~\ref{fig:pss3d}\,(c)). 
In phase-space there is a single large chaotic attractor left and thus, the `skeletons'
of the formerly transporting attractors have completely disappeared.

\section{Conclusion}\label{sect:summary}
We have investigated the dynamics of two coupled driven and damped particles whose individual motions sees
one particle evolving on a periodic attractor, and the other evolving on a strange chaotic attractor.
In the low coupling regime it is the particle evolving on the periodic attractor that has most influence
with respect to the system dynamics. Here a directed current can be seen. However, with increasing strength of
coupling
both attractors play a key role in the overall dynamics. Indeed, it has been demonstrated that with increasing
coupling strength the attractors can merge thus allowing trajectories to explore wider regions of phase-space.
We have also shown that the chaos-hyperchaos transition, appearing at a critical coupling strength,
marks the point at which a directed current ceases to exist. 
the frequency and amplitude of the
In addition these windows are quite often too narrow to allow for a 
We performed further studies combining the parameter values from 
different windows of regular and chaotic behaviour 
occurring in the bifurcation diagrams of the single (uncoupled) system \cite{single}. 
In general it turns out that a merging crisis is always a precursor to a transition to hyperchaos 
on the route to vanishing current.  

Relating these results to a previous study \cite{chaos} where both particles evolve (individually) on strange
attractors, 
we find that the system under investigation in this
paper
promotes directed transport in a regime of much weaker coupling. However, increased coupling can produce
contrasting results for both systems. For the system described above, an increased coupling yields global
chaos resulting in zero current, while in \cite{chaos} a directed current is only possible with an increased
coupling such that the chaos is suppressed.

\begin{acknowledgements}
Numerical computations were done on the Sciama High Performance Compute (HPC) cluster which is supported by 
the ICG, SEPNet and the University of Portsmouth.
\end{acknowledgements}

\bibliographystyle{abbrv}

\end{document}